\documentclass[sigconf, nonacm]{acmart}

\newcommand\vldbyear{2025}
\newcommand\vldbworkshop{LLM+Graph}
\newcommand\vldbauthors{\authors}
\newcommand\vldbtitle{\shorttitle} 
\newcommand\vldbavailabilityurl{}
\newcommand\vldbpagestyle{empty}

\ccsdesc[500]{Computing methodologies~Spatial and physical reasoning}


\usepackage[]{svg}
\usepackage{multirow}
\usepackage{caption}
\usepackage{subcaption}
\usepackage{algorithm}
\usepackage{algpseudocode}
\usepackage{afterpage}
\usepackage{tablefootnote}
\usepackage{appendix}
\usepackage{float}
\usepackage{gensymb}
\usepackage{natbib}

\usepackage{xcolor}

\definecolor{codegreen}{rgb}{0,0.6,0}
\definecolor{codegray}{rgb}{0.5,0.5,0.5}
\definecolor{codepurple}{rgb}{0.58,0,0.82}
\definecolor{backcolour}{rgb}{0.95,0.95,0.92}
\usepackage{listings}
\lstdefinestyle{mystyle}{
    backgroundcolor=\color{backcolour},   
    commentstyle=\color{codegreen},
    keywordstyle=\color{magenta},
    numberstyle=\tiny\color{codegray},
    stringstyle=\color{codepurple},
    basicstyle=\ttfamily\footnotesize,
    breakatwhitespace=false,         
    breaklines=true,                 
    captionpos=b,                    
    keepspaces=true,                 
    numbers=left,                    
    numbersep=5pt,                  
    showspaces=false,                
    showstringspaces=false,
    showtabs=false,                  
    tabsize=2
}
\newbool{showcomments}

    \booltrue{showcomments}

    \newcommand{\pinaforecomment}[4]{%
    \ifbool{showcomments}{%
        \colorbox{#1}{\textcolor{#4}{\parbox{.8\linewidth}{#2: #3}}}%
    }{}%
}

\usepackage{acronym}

\newacro{AI}{Artificial Intelligence}
\newacro{BPM}{Bipartite Matching}
\newacro{CNN}{Convolutional Neural Network}
\newacro{CSP}{Constraint Satisfaction Problem}
\newacro{DFS}{Depth First Search}
\newacro{FAISS}{Facebook Artificial Intelligence Similarity Search}
\newacro{GPS}{Global Positioning System}
\newacro{HGNN}{Heterogeneous Graph Neural Network}
\newacro{KG}{Knowledge Graph}
\newacro{LLM}{Large Language Model}
\newacro{LoRA}{Low-Rank Adaption}
\newacro{m-CK}{m-Closest Keywords}
\newacro{MSE}{Mean Squared Error}
\newacro{NL2PQ}{Natural Language to Pictorial Query}
\newacro{NLP}{Natural Language Processing}
\newacro{OOV}{Out-Of-Vocabulary}
\newacro{OSM}{Open Streetmap}
\newacro{POI}{Point of Interest}
\newacro{RAG}{Retrieval Augmented Generation}
\newacro{RDF}{Resource Description Framework}
\newacro{SGM}{Subgraph Matching}
\newacro{SKQ}{Spatial Keyword Query}
\newacro{SPARQL}{SPARQL Protocol and RDF Query Language}
\newacro{SQL}{Structured Query Language}
\newacro{t-SNE}{t-Stochastic Neighbor Embedding}

\begin{document}


\graphicspath{ {figures/}{auto_commit_fig/}{auto_fig/} }

\newcommand{\latexfile}[1]{\input{sections/#1}}

\newcommand{\osullikomment}[1]{\pinaforecomment{green}{Kent}{#1}{black}}
\newcommand{\nrscomment}[1]{\pinaforecomment{violet}{Nicole}{#1}{white}}


\title{Graph-Enhanced Large Language Models for Spatial Search}

\author{Nicole R. Schneider}
\email{nsch@umd.edu}
\affiliation{%
  \institution{University of Maryland}
  \city{College Park}
  \country{USA}
}

\author{Kent O'Sullivan}
\email{kosu0918@uni.sydney.edu.au}
\affiliation{%
  \institution{University of Sydney}
  \city{Sydney}
  \country{Australia}
}

\author{Hanan Samet}
\email{hjs@cs.umd.edu}
\affiliation{%
  \institution{University of Maryland}
  \city{College Park}
  \country{USA}
}

\begin{abstract}
There have been many recent improvements in the ability of Large Language Models (LLMs) to perform complex tasks and answer domain-specific questions through techniques like Retrieval Augmented Generation (RAG).
However, reasoning abilities of LLMs, including spatial reasoning abilities, are still lacking.
Spatial reasoning is a key component required to answer questions in a variety of domains that are grounded in the physical world, including urban planning, civil engineering, travel, and many others.
To advance the development of LLMs and facilitate an impact in these domains, new research techniques must be developed to enable LLMs to reason over spatial data, which is commonly stored in the form of a graph.
In this paper we outline the challenges associated with spatial reasoning through LLMs and envision a future in which search engines integrate with LLMs to answer complex spatial questions through graph-enhanced reasoning.

\end{abstract}

\begin{teaserfigure}
    \includegraphics[width=0.9\linewidth]{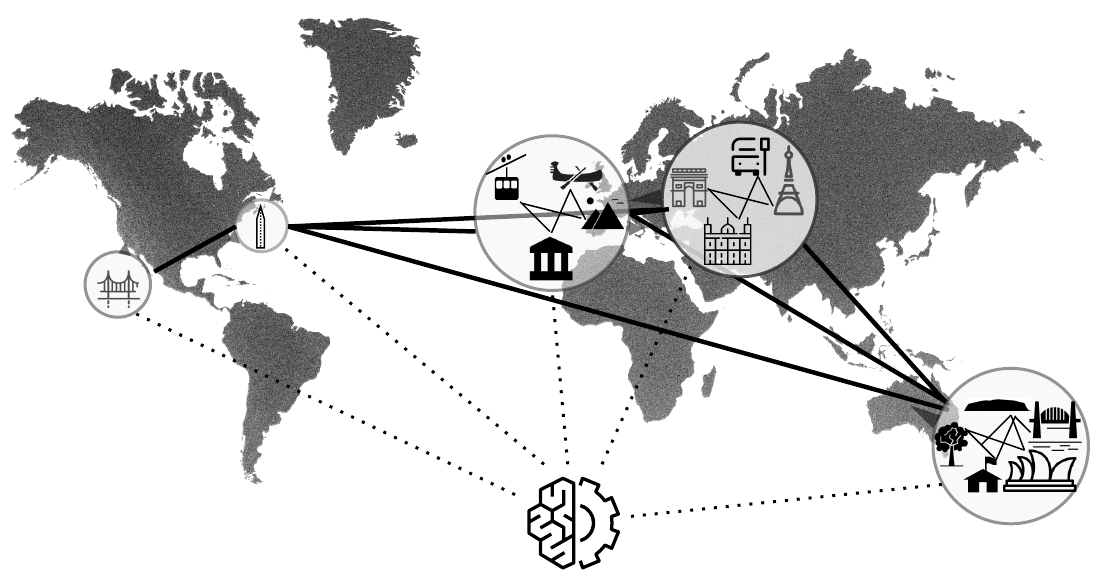}
    \centering
    \caption[width=0.7\linewidth]{A world model for LLMs.}
    \Description{Map of the world synthesized as a spatial graph with a model attached}
    \label{fig:teaser}
\end{teaserfigure}

\maketitle

\pagestyle{\vldbpagestyle}
\begingroup\small\noindent\raggedright\textbf{VLDB Workshop Reference Format:}\\
\vldbauthors. \vldbtitle. VLDB \vldbyear\ Workshop: \vldbworkshop.\\ 
\endgroup
\begingroup
\renewcommand\thefootnote{}\footnote{\noindent
This work is licensed under the Creative Commons BY-NC-ND 4.0 International License. Visit \url{https://creativecommons.org/licenses/by-nc-nd/4.0/} to view a copy of this license. For any use beyond those covered by this license, obtain permission by emailing \href{mailto:info@vldb.org}{info@vldb.org}. Copyright is held by the owner/author(s). Publication rights licensed to the VLDB Endowment. \\
\raggedright Proceedings of the VLDB Endowment. 
ISSN 2150-8097. \\
}\addtocounter{footnote}{-1}\endgroup

\ifdefempty{\vldbavailabilityurl}{}{
\vspace{.3cm}
\begingroup\small\noindent\raggedright\textbf{VLDB Workshop Artifact Availability:}\\
The source code, data, and/or other artifacts have been made available at \url{\vldbavailabilityurl}.
\endgroup
}

\begingroup
\renewcommand{\arraystretch}{1.6}
\setlength{\tabcolsep}{5pt}
\begin{table*}[ht]
    \centering
    \normalsize
    \begin{tabular}
    {p{60pt} p{60pt} p{60pt} p{190pt} r}
        \hline
        \hline
        \textbf{Relation} & \textbf{Domain} & \textbf{Task} & \textbf{Potential Use Case or Application} & \textbf{Examples} \\
        \hline
        Metric      & Realty & Housing Search          & ``Which homes are within 10km of a school and more than 5km from a highway?'' & \cite{Chen2019} \\
        ~           & Travel & Itinerary \mbox{Planning}      & ``Find a (library, museum, restaurant) or a (museum, library, restaurant) where the distance from one place to the next is less than 400m.'' & \cite{Fang2019} \\
        \hline
        Topological & Civil \mbox{Engineering} & Road \mbox{Network} Alignment  & ``Given a sketch drawing of street intersections and highways, which cities are a match?'' & \cite{Schwering2014} \\
        ~           & Agriculture & Multi-Scale Data Fusion & ``Which land cover datasets at different scales are a spatial match to this one?'' & \mbox{\cite{Huang2023}} \\
        \hline
        Directional & Event \mbox{Planning} & Landmark Search         & ``Which restaurant has an ocean to the East, a parking lot to the North, and a golf course to the south?'' & \cite{Osul2023b} \\
        \hline
        \hline
    \end{tabular}
    \caption{Application domains of spatial-pattern search where graph-enhanced LLMs can provide potential benefit.}
    \label{tab:11_relation_examples}
\end{table*}
\endgroup

\section{Introduction}
\label{section:introduction}

There have been many recent improvements in the ability of \acp{LLM} to perform complex tasks and answer domain-specific questions through \ac{RAG}~\cite{Lewis2020} and other techniques.
\ac{LLM}s have been used with great success in areas that involve human abstractions, like writing code~\cite{Liu2023b}, summarizing and translating text~\cite{Laban2023,Schneider2024c}, and performing math calculations~\cite{Didolkar2024}.
However, \ac{LLM}s struggle to perform well when a task requires spatial or temporal reasoning because they lack a coherent, consistent world model, like the ones designed in reinforcement learning settings~\cite{Ha2018,Backlund2025}.
Further, in many domains, search queries are naturally expressed through an abstract representation of spatial constraints as opposed to the keyword search terms that most search engines and \ac{LLM}s are accustomed to handling.
In urban planning, road network alignment is used to identify existing places with similar road layouts to a query location or proposed road plan~\cite{Schwering2014}
In the travel domain, a user may recall a specific \ac{POI} by the spatial configuration of the landmarks around it, without remembering the exact address or name of the place~\cite{Osul2023,Osul2023b}.
In event planning, a user may not have a specific place in mind, but instead have some spatial criteria they seek to satisfy, such as finding the set of restaurants in Washington D.C. that are located next to a park and within half a kilometer of a metro stop for easy accessibility.
In these examples, as well as other use cases outlined in Table \ref{tab:11_relation_examples}, an abstract spatial pattern is the most natural expression of the user's search intention.

To advance the development of LLMs to aid in answering spatial questions like the ones we describe, new research techniques must be developed to enable \ac{LLM}s to reason over spatial data.
Such a development would require LLMs to internalize an accurate spatial model of the world that would allow them to compare places based on their physical locations and infer spatial relationships through transitivity (Figure \ref{fig:teaser}).
Although recent work has shown that \ac{LLM}s possess a rough model of the world intrinsically through training~\cite{Bhandari2023, Qi2023}, they lack knowledge of less common places and cannot generalize or infer spatial relationships, such as the distances between places~\cite{Osullivan2024}.

The need for spatially-aware search is not presently filled by mainstream search and mapping platforms, leaving a gap that LLM-aided spatial search may be able to address.
Current search engines and mapping platforms like Google Maps~\footnote{\href{https://developers.google.com/maps/documentation/places/web-service/search-text}{https://developers.google.com/maps/documentation/places/web-service/search-text}} are designed for keyword search, not spatial search.
That is, they can support \mbox{top-k} keyword queries that seek co-located objects relevant to query keywords, but they cannot support queries involving spatial relationships between keyword entities~\cite{Chen2022}.
Taking the previous example of searching for restaurants located next to a park and within half a kilometer of a metro stop, \mbox{top-k} search would return any restaurants, parks, and metro stops in the area of interest, ignoring the desired distance and topological constraints.
Even spatial group keyword search, which is more complex than \mbox{top-k} search, would find groups of those entities in a minimum enclosing circle, but still would not satisfy the topological constraint (restaurant adjacent to park)~\cite{Cao2015,Mahmood2017,Mahmood2019}.
Instead, to fulfill search intentions expressed by abstract spatial patterns, an NP-hard form of search called spatial pattern matching, or pattern-based spatial search, is needed~\cite{Fang2019}.

Pattern-based spatial search seeks to match query keywords and spatial constraints to the data entities that satisfy them. 
Query constraints can capture metric, topological, and directional relationships between entities, based on how they are positioned in space.
These constraints can describe explicit relationships between pairs of entities (i.e. X north of Y), in the form of edges in a graph, or implicit relationships, such as through a pictorial representation or sketch map query~\cite{Folkers2000,Schwering2014}.
Spatial pattern matching supports queries with arbitrary spatial constraints, making it useful for the spatial search scenarios we describe, but it has several challenges that inhibit its use in search engines and \ac{LLM}s.

In this paper, we first outline the challenges of reasoning about spatial relationships, then describe research directions that can be undertaken to address these challenges and enable pattern-based spatial search using \ac{LLM}s.
Specifically, we highlight:

\begin{enumerate}
    \item The computational challenges associated with resolving spatial queries, 
    \item The heterogeneous nature of spatial entities and relations, which makes reasoning over them difficult, and
    \item The graph or pictorial query formats, which are not naturally compatible with text-centric search engines and \ac{LLM}s.
\end{enumerate}

\noindent After describing these challenges, we present a vision for how graph-enhanced \ac{LLM}s can play a role in enabling spatial search, thus equipping users with a form of search not currently available with mainstream search engines.

\subsection{Vision}
We envision a future where search engines integrate with \ac{LLM}s to answer complex spatial questions using graph-enhanced reasoning.
Realizing this vision will make pattern-based spatial search accessible for widespread use in the applications we describe.
To accomplish accessible pattern-based spatial search, each of the challenges outlined above must be addressed; computational complexity, heterogeneity of entity/relation types, and custom query formats. 
In this vision paper, we describe several research directions that can be undertaken to address these challenges by leveraging graph-enhanced \ac{LLM}s.
We propose that \ac{LLM}s can be used to interpret spatial questions and formulate queries that are answerable by tailored spatial reasoning models, which can use spatial databases and graphs to find and rank results that answer the question.
By tailoring specific learning objectives or data stores to different categories of spatial information, the key spatial properties, like directional transitivity, can be learned and then leveraged in an ensemble approach to spatial reasoning.
We further envision the use of \ac{NLP} techniques to interpret spatial patterns described in natural language and convert them into graph or pictorial queries, which can be reasoned about, enabling seamless integration with the pre-existing text interfaces used by \ac{LLM}s and by most search interfaces.

\subsection{Contribution}
Our primary contributions are an analysis of the present challenges preventing the broad use of spatial pattern search and a road map of a future where \ac{LLM}s provide a means to solve those challenges.
We identify real-time search scenarios that would benefit from the adoption of pattern-based spatial search and describe several research directions that may lead graph-enhanced \ac{LLM}s to enable fast, robust approximate pattern-based spatial search.
We further highlight the need for a benchmark to measure the accuracy and scalability of pattern-based spatial search approaches that rely on \ac{LLM}s compared to traditional spatial pattern matching methods that have correctness guarantees.
We see this work as a natural extension to the current research efforts to achieve reasoning through \ac{LLM}s using techniques like \ac{RAG}.
We hope that by identifying interesting and worthwhile open problems in graph-enhanced \ac{LLM} spatial reasoning, this work will spark further efforts to equip \ac{LLM}s with new abilities that make them better suited to serve everyday users in a variety of applications.

\section{Challenges of Spatial Search}
\label{section:challenges}

Spatial search has a variety of challenges that make it difficult to support by current search engines and mapping platforms.

\subsection{Computational Cost}
Geospatial data is naturally stored as a graph, with edge attributes explicitly capturing the relationships between places (nodes).
Spatial pattern matching over geospatial graphs is generally regarded as intractable for large search spaces~\cite{Schneider2024,Schneider2024b}.
The computational complexity stems from the quadratic number of relations that can be defined on a set of entities, which makes the process of matching query and data patterns NP-hard~\cite{Fang2019}.
Unlike less expressive forms of spatial search, such as spatial group keyword search, which is also NP-hard, there are no approximate methods that reduce the complexity of spatial pattern matching broadly.
Since search engines and mapping applications typically support real-time search, current spatial pattern matching methods are too computationally burdensome to incorporate into their pipelines.

\subsection{Heterogeneous Spatial Relationships}
Spatial patterns can be defined by metric, topological, and distance relationships between entities.
These relations can be quantitative or qualitative, and they can be defined over point, line, and/or region data.
With many possible combinations to account for, it is difficult to reason about all of them with one generic approach.
Instead there are many different methods that work on different subsets of data and relation types.
For example, tailored spatial pattern matching systems like 
Spacekey~\cite{Fang2019} and 
SketchMapia~\cite{Schwering2014} 
have been developed to handle use cases like housing search (e.g. ``Find all the houses in an area that are within 10km of a school and greater than 5km from any highway.'') or
road network alignment (e.g. ``Given a sketch drawing of street intersections and highways, which cities are a match?'').
While these systems can handle some pattern-based spatial search queries, they are scoped to single use cases that involve only a subset of possible entity and relation types.
Hence the methods they use are not generalizable.

\subsection{Graph Query Input}
Finally, spatial pattern matching typically requires a query pattern, given in the form of a sketch map or pictorial query that is drawn or constructed on a digital canvas, by dragging and dropping entities and arranging them in the spatial layout of interest~\cite{Fang2018,Osul2023}.
This query format offers exceptional expressivity to capture the user's search intention, and it aligns with the intuitive visual nature of spatial pattern queries, mimicking the process of hand-drawing a sketch map on paper or imagining oneself at a location, capturing spatial relationships between objects in view~\cite{Schwering2014}.
However, accepting queries in a pictorial or sketch map format necessitates having a custom interface with a mechanism to construct the graph or pictorial representation, including by selecting, dragging, and dropping entities, and possibly labeling edges with relationship constraints, such as ``less than 3 km.''
Such an interface requires capabilities beyond the standard text input used by most \ac{LLM}s and search engines, making it structurally difficult to accomplish spatial search through existing platforms.

\begin{figure}[ht]
    \includegraphics[width=\linewidth]{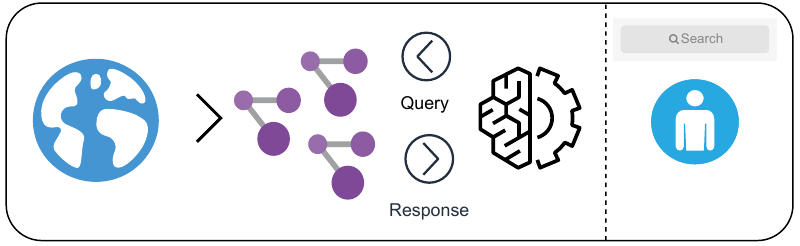}
    \centering
    \caption[width=0.7\linewidth]{Vision for graph-enhanced search using an LLM. A user interacts with the LLM, which interprets the query and leverages appropriate spatial models and graphs to answer the search question.}
    \label{fig:pipeline}
    \Description{Pipeline showing abstract concept of user with search function accessing LLM which queries graphs about the world to get a response}
\end{figure}

\begin{figure*}[ht]
    \includegraphics[width=\linewidth]{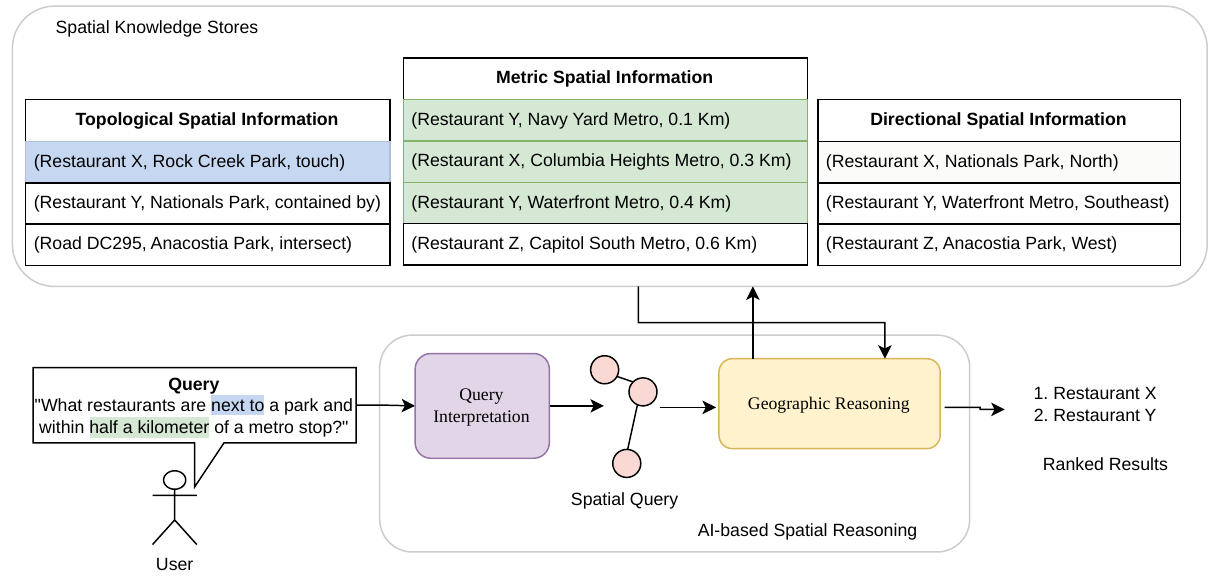}
    \centering
    \caption[width=0.7\linewidth]{Vision for spatial search using graph-enhanced \ac{LLM}s. User inputs a spatial query, which is interpreted and converted into a graph query. Geographic reasoning agent or module uses query to select geospatial knowledge stored in graph(s) and perform graph reasoning to resolve the query. Ranked results are outputted.}
    \Description{Diagram showing use of metric, topological, and directional data stores to perform geographic reasoning}
    \label{fig:storage}
\end{figure*}

\section{Vision for Spatially-Aware LLMs}
\label{section:vision}

We envision future research using graph-enhanced \ac{LLM}s to address the challenges causing the current gap in pattern-based spatial search (Figure \ref{fig:pipeline}).
Specifically, we envision the computational challenges being addressed through approximation, the heterogeneous nature of spatial entities and relations being addressed through an ensemble of methods, learning objectives, and data sources, and the graph query input challenge being addressed through natural language to graph conversion using pre-trained language models.
Below we describe each of these ideas in greater depth.

\subsection{Computational Efficiency Through Approximation}

To alleviate the computational challenges associated with resolving spatial queries that require searching for patterns in a large quantity of spatial entities, we envision achieving efficient, real time search through approximation.
Traditional spatial pattern matching methods typically perform exact search, often using pruning and other techniques to reduce some of the computational burden.
A few approaches perform approximate search, but these are typically limited in expressivity, supporting only point entities and metric relations~\cite{Chen2013}, or localizing the search to the entities that are relevant for a particular problem space, like roads and city blocks for the road network alignment task~\cite{Schwering2014}.
We envision leveraging graph-enhanced \ac{LLM}s that can access reference spatial information to perform flexible, approximate spatial pattern search across a variety of domains.
Using \ac{LLM}s would avoid the NP-hard complexity of exact spatial pattern search, while maintaining flexibility, which is critical to support multiple domains and deal with noisy data tags, which are commonplace in spatial data.

To ensure approximate \ac{LLM}-based methods achieve reasonable accuracy, a spatial pattern matching benchmark should be developed.
State of the art spatial pattern matching approaches typically use pruning techniques, binning, localization, and approximate methods to search through candidate entities efficiently. 
However, the accuracy and efficiency of these approaches can depend on the size of the localized region or pruned entity set, which may vary significantly based on the query, dataset, and method being used.
Presently, empirical testing is done using ad-hoc datasets which vary from paper to paper.
Without a common benchmark or leader-board, spatial pattern matching methods cannot be rigorously evaluated with respect to one another and the trade-off between accuracy and efficiency of different techniques cannot be fully understood~\cite{Osullivan2024}.
To elicit worst-case behavior in traditional methods for a fair comparison, we envision a spatial pattern matching benchmark that includes complex datasets and large query patterns covering the major types of spatial entities and relations.
Further, such a benchmark should vary query and dataset characteristics that may influence the size of the pruned candidate set for some methods, which can significantly impact runtime.

Of particular importance, based on initial exploration, is the size of the query pattern, which has a large impact on performance.
We further observed that most of the current spatial pattern matching approaches test their method on ad-hoc datasets containing query patterns of no more than six entities~\cite{Chen2019,Minervino2023}.
Based on the theoretical complexities derived in \citeauthor{Schneider2024b}, spatial pattern search over larger query patterns or complex questions with many spatial constraints remains an open problem, with most existing methods having a complexity that is exponential in the size of the query~\cite{Schneider2024b}.
As a result, including large query patterns in a benchmark will be critical to evaluating existing approaches and weighing the benefits of approximate methods, like graph-based \ac{LLM} methods, that likely scale better, at the cost of search precision and recall.

\subsection{Reasoning Over Heterogeneous Spatial Relationships with Graph-Enhanced LLMs}
To address the heterogeneity in spatial entities and relations that can appear in spatial reasoning questions, we suggest using an ensemble of models or reference graph databases, each of which is tailored to a certain type of spatial relation (Figure \ref{fig:storage}).
For instance, following \citeauthor{Schneider2025b}, one graph may include distances on the edges, making it useful for metric spatial reasoning questions~\cite{Schneider2025b,Schneider2025}, and another may include nodes representing points, lines, and regions with qualitative relationships between them, making it useful for topological spatial reasoning.
By doing retrieval from each of these separate data stores, or by developing expert models that are trained to do different types of spatial reasoning, the results can be combined to achieve rigorous reasoning over the heterogeneous set of possible spatial relationships.

A spatial reasoning mechanism can be developed in a variety of ways, including by using an \ac{HGNN}~\cite{Zhang2019} or Heterogeneous Graph Transformer~\cite{Hu2020} architecture, where the learning objective complements the goal of spatial reasoning.
Instead of the masked language model or next token prediction objectives that enable \ac{LLM}s to learn grammatical structure of language, a spatial learning objective, like a distance reconstruction learning objective, could be developed.
For example, to accomplish metric spatial reasoning, random walks could be taken through a fully connected graph of places and distances between them, where some edges and nodes are masked. 
Then, the model could reconstruct the full walk, using the masked version, allowing the model to learn to predict how far apart different places are to each other, by identifying them in a reference graph and aligning the masked walk to its corresponding reference subgraph.

A topological spatial reasoning model would need to encode region and line data to enable topological reasoning.
These types of geographic data are usually stored in trees, so a learning objective could seek to reconstruct the topological relationships by traversing a masked region quadtree.
Since there are 9 distinct topological operators~\cite{Carniel2023}, one approach may be to develop one topological sub-model per operator, and have a topological relationship identifier to determine the appropriate model to use given clues in the query, such as the word ``contains'' or ``intersects.''
Tailored approaches can also be developed for directional spatial pattern matching, using graphs to encode the critical spatial properties needed to answer the questions, and developing models capable of reasoning over that information.

\subsection{Converting Natural Language into Spatial Graph Queries Using Language Models}

To bridge the gap between pattern-based spatial search and traditional search engines and 
platforms that require textual input, we draw inspiration from the Natural Language to Structured Query Language (NL2SQL) literature which uses \ac{LLM}s to generate structured queries based on natural language input~\cite{Kim2020,Brunner2021,Fu2023}.
We envision an \ac{LLM}-based conversion process that turns natural language text into a pictorial spatial pattern representation, enabling pattern-based spatial search via natural language input.
While previous work has investigated methods for semantic parsing of spatial information from natural language for robotics and other applications~\cite{Dukes2014,Kim2024}, no studies have tackled the problem of generating pictorial or graph queries for spatial pattern matching given text input.
Moreover, many semantic parsing methods for extracting spatial relations are brittle, making them unsuitable for arbitrary text input from users describing spatial patterns for search.
Instead, to enable spatial pattern matching over natural language input, an LLM could be used to interpret a user's text input and construct a query or series of queries over graph data stores containing spatial entities and relation constraints.

The challenge of applying \ac{LLM}s to tasks involving geospatial reasoning is further complicated by the lack of representation of lesser-known places in training data.
Many obscure places have long, complex names that are out-of-vocabulary for \ac{LLM}s, making it even harder for \ac{LLM}s to associate them with their physical location in the world and reason about them.
Given the vast array of possible places that can be included in a question requiring spatial reasoning, a basic `world model' is needed to complement the existing linguistic knowledge held in \ac{LLM}s, with the ability to pull more obscure places from a data store, rather than relying on trained knowledge of every place on earth.

\section{Conclusion}
\label{section:conclusion}
\normalsize

Spatial reasoning is a key component required to answer many questions.
New research techniques must be developed to enable LLMs to reason over spatial data, which is commonly stored in the form of a graph.
We outline several challenges associated with spatial reasoning through LLMs, including computational complexity, heterogeneity of spatial entities and relations, and non-textual query input format.
To address these issues we envision a future in which search engines integrate with LLMs that use graph-enhanced reasoning to answer complex spatial questions.
We hope that by identifying interesting and worthwhile open problems in graph-enhanced \ac{LLM} spatial reasoning, this work will spark further efforts to equip \ac{LLM}s with new abilities that make them better suited to serve everyday users in a variety of applications.

\begin{acks}
This work was sponsored in part by the \grantsponsor{1}{NSF}{} under Grants \grantnum{1}{IIS-18-16889}, \grantnum{1}{IIS-20-41415}, and \grantnum{1}{IIS-21-14451}.
\end{acks}

\bibliographystyle{style/ACM-Reference-Format.bst}
\bibliography{main.bib} \label{bibliography}

\end{document}